\documentclass[12pt]{article}
\usepackage{amsmath}
\usepackage{amssymb}
\begin{document}
\numberwithin{equation}{section}
\numberwithin{table}{section}

\def\AEF{A.E. Faraggi}
\def\NPB#1#2#3{{\it Nucl.\ Phys.}\/ {\bf B#1} (#2) #3}
\def\PLB#1#2#3{{\it Phys.\ Lett.}\/ {\bf B#1} (#2) #3}
\def\PRD#1#2#3{{\it Phys.\ Rev.}\/ {\bf D#1} (#2) #3}
\def\PRL#1#2#3{{\it Phys.\ Rev.\ Lett.}\/ {\bf #1} (#2) #3}
\def\PRT#1#2#3{{\it Phys.\ Rep.}\/ {\bf#1} (#2) #3}
\def\MODA#1#2#3{{\it Mod.\ Phys.\ Lett.}\/ {\bf A#1} (#2) #3}
\def\IJMP#1#2#3{{\it Int.\ J.\ Mod.\ Phys.}\/ {\bf A#1} (#2) #3}
\def\nuvc#1#2#3{{\it Nuovo Cimento}\/ {\bf #1A} (#2) #3}
\def\etal{{\it et al}\ }

\begin{table}[t]
  \begin{flushright}
     OUTP-03-02P    \\ 
     hep-th/0301147 \\
     January 2003   
  \end{flushright}  
\end{table}

\title{\bf Yukawa couplings \\in $SO(10)$ heterotic M-theory vacua}

\author{Alon E. Faraggi\thanks{faraggi@thphys.ox.ac.uk} \ 
and Richard S. Garavuso\thanks{garavuso@thphys.ox.ac.uk} \\
\emph{\normalsize Theoretical Physics Department, University of Oxford} \\
\emph{\normalsize Oxford \ OX1 3NP, UK}}

\date{ }

\maketitle
\thispagestyle{empty}

\begin{abstract}
We demonstrate the existence of a class of $ \mathcal{N} = 1 $
supersymmetric nonperturbative vacua of Ho\v{r}ava-Witten M-theory
compactified on a torus fibered Calabi-Yau
3-fold $ Z $ with first homotopy group $ \pi_{1} (Z) = \mathbb{Z}_{2} $,
having the following properties: 1) $ SO(10) $ grand unification group,
2) net number of three generations of chiral fermions in the
observable sector, and 3) potentially viable matter Yukawa couplings. 
These vacua correspond to semistable holomorphic vector bundles 
$ V_{Z} $ over $ Z $ having structure group $ SU(4)_{\mathbb{C}} $, and 
generically contain M5-branes in the bulk space. The nontrivial
first homotopy group allows Wilson line breaking of the $ SO(10) $
symmetry. Additionally, we propose how the 11-dimensional
Ho\v{r}ava-Witten M-theory framework may be used to extend
the perturbative calculation of the top quark Yukawa coupling in the
realistic free-fermionic models to the nonperturbative regime. 
The basic argument being that the relevant coupling couples
twisted-twisted-untwisted states and can be 
calculated at the level of the $ Z_{2} \times Z_{2} $
orbifold without resorting to the full three generation models.

\end{abstract}

\pagebreak

\tableofcontents

\section{Introduction} 
\setcounter{page}{1}

The five self-consistent 10-dimensional superstring theories are different
vacua of a single underlying 11-dimensional quantum theory,
\emph{M-theory}, which has 11-dimensional supergravity as its low
energy limit~\cite{M-theory}.  While the complete formulation of M-theory
is not known, a web of perturbative and nonperturbative dualities 
has been established which connects the different M-theory limits.  These
dualities provide insight into the nonperturbative behavior of the
superstring theories.

One such duality, proposed by Ho\v{r}ava and Witten,
connects M-theory compactified on an orbifold 
$ S^{1}/\mathbb{Z}_{2} $ with the strong coupling limit of the $ E_{8}
\times E_{8} $ heterotic string~\cite{HW}. A class of 
$ E_{8} \times E_{8} $ heterotic string models are the
\emph{realistic free-fermionic models}~\cite{rffm}.
A remarkable achievement of the realistic free-fermionic models is their
successful prediction of the top quark mass~\cite{Far:Top}.

The first goal of this paper is to demonstrate the existence of a class of 
$ \mathcal{N} = 1 $ supersymmetric nonperturbative vacua of
Ho\v{r}ava-Witten M-theory compactified on a torus fibered Calabi-Yau
3-fold $ Z $ with first homotopy group 
$ \pi_{1}(Z) = \mathbb{Z}_{2} $, having the following properties:
\begin{enumerate}

\item $ SO(10) $ grand unification group.

\item Net number of generations $ N_{gen} = 3 $ of chiral fermions in the
observable sector.

\item Potentially viable matter Yukawa couplings.

\end{enumerate}
The nontrivial first homotopy group allows Wilson line breaking of the   
grand unification group. Our second goal is to discuss how the
11-dimensional Ho\v{r}ava-Witten M-theory framework may be used to extend
the perturbative calculation of the top quark Yukawa coupling in the
realistic free-fermionic models to the nonperturbative regime.

The tools needed to achieve the above goals have been recently developed.
Donagi, Ovrut, Pantev and Waldram~\cite{DOPW,DOPW2} presented rules for
constructing a class of $ \mathcal{N}=1 $ supersymmetric nonperturbative
vacua of Ho\v{r}ava-Witten
M-theory compactified on a torus fibered Calabi-Yau 3-fold $ Z $ with
first homotopy group $ \pi_{1}(Z) = \mathbb{Z}_{2} $, having grand
unification group $ H = E_{6} $ or $ H = SU(5) $ and net number of 
generations $ N_{gen} = 3 $ of chiral fermions in the observable sector.
The case with grand unification group $ H = SO(10) $ 
was studied in~\cite{FGI1}, where the overlap with the free-fermionic
models was discussed.  The vacua with $ H = E_{6} $, $ SO(10) $, or
$ SU(5) $ correspond to semistable holomorphic vector bundles $ V_{Z} $
over $ Z $ having structure group 
$ G_{\mathbb{C}} = SU(3)_{\mathbb{C}} $, $ SU(4)_{\mathbb{C}} $, or 
$ SU(5)_{\mathbb{C}} $, respectively, and generically contain M5-branes in
the bulk space. Arnowitt and Dutta~\cite{AD1} argue that
phenomenologically viable matter Yukawa couplings can be obtained by
requiring vanishing instanton charges on the observable orbifold fixed
plane and clustering of the M5-branes near the hidden orbifold fixed
plane, and provide an explicit $ H = SU(5) $ example.

To aid in achieving our first goal of obtaining vacua with $ H = SO(10) $,
$ N_{gen} = 3 $, and potentially viable matter Yukawa couplings, we 
combine the $ G_{\mathbb{C}} = SU(4)_{\mathbb{C}} $ rules discussed 
in~\cite{FGI1} with the constraint of vanishing instanton charges on
the observable orbifold fixed plane. Instead of restricting
ourselves to a set of sufficient (but not necessary) constraints on the
vector bundles (as was done in~\cite{DOPW,DOPW2,FGI1,AD1}), 
we consider the most general case.  Indeed, the vacua obtained in
Section~\ref{DPez} require this generalization. 

The key to achieving our second goal is to utilize the correspondence
between the free-fermionic models and  
$ \mathbb{Z}_2\times \mathbb{Z}_2 $
orbifold compactification~\cite{foc}. In the free fermionic models
the top quark mass term arises from a twisted--twisted--untwisted
Yukawa coupling, at a fixed point in the moduli space.
We can calculate this coupling in the three generation model,
or at the level of the (51,3) or (27,3) $Z_2\times Z_2$ orbifold.
While we do not know the precise
geometrical realization of the three generation models, the geometry of
the $ \mathbb{Z}_2\times \mathbb{Z}_2 $ orbifold is more readily identified.
Furthermore, the calculation can be done as either a 
$ \mathbf{16} \cdot \mathbf{16} \cdot \mathbf{10}$ $SO(10)$ coupling,
or a $ \mathbf{27}^{3} $ $E_{6}$ coupling.
At the free-fermionic point in the moduli space 
the numerical results will be identical.
Thus, to extend the calculation of the top quark Yukawa
coupling in the realistic free-fermionic models to the nonperturbative
regime, one can compactify Ho\v{r}ava-Witten M-theory on a Calabi-Yau
3-fold which corresponds to the $ \mathbb{Z}_2\times \mathbb{Z}_2 $
orbifold.  One can choose $ SU(n)_{\mathbb{C}} $ vector bundles with 
$ n = 3 $ or $n=4$, corresponding to the $E_{6}$ or $SO(10)$ grand 
unification group, respectively. The nonperturbative top quark Yukawa
coupling at the grand unification scale is then computed, at least in
principle, using~(\ref{Yukawa}).  We note that the compactification on the
Calabi-Yau 3-fold corresponding to the  
$ \mathbb{Z}_2\times \mathbb{Z}_2 $ orbifold will lead to 
$N_{gen} \neq 3$.

We are thus led to present rules for arbitrary $ N_{gen} $, allowing
general $ SU(n)_{\mathbb{C}} $ vector bundles with $ n = 3 $, 4, or 5
corresponding to grand unification group $ E_{6} $, $ SO(10) $, or 
$ SU(5) $, respectively.  To obtain potentially viable Yukawa couplings,
we require vanishing instanton charges on the observable orbifold fixed
plane.  This further restricts the allowed vector bundles. 

This paper is organized as follows.  In Section~\ref{HorWit}, we briefly
review Ho\v{r}ava-Witten M-theory, some results of its compactification to
four dimensions and the associated 4-dimensional low energy effective
theory. In Section~\ref{Rules}, we present the rules as discussed above.
Section~\ref{Hirz} demonstrates that torus fibered Calabi-Yau 3-folds with
$ \pi_{1}(Z) = \mathbb{Z}_{2} $ and a Hirzebruch base surface do not admit
the $ H = SO(10) $, $ N_{gen} = 3 $ vacua with potentially viable
matter Yukawa couplings that we seek.  In contrast, Section~\ref{DPez}
demonstrates the existence of such vacua with a del Pezzo $ dP_{7} $ base.
In Section~\ref{FreeFerm}, we discuss the extension of the top quark
Yukawa coupling calculation in the realistic free-fermionic
models to the nonperturbative regime, and explain why
modifications to the rules of Section~\ref{Rules} may be required for a
detailed analysis.  Section~\ref{Conc} summarizes our conclusions.

\section{\label{HorWit}Review of Horava-Witten M-theory}

Ho\v{r}ava and Witten proposed that M-theory compactified
on an orbifold $ S^{1}/\mathbb{Z}_{2} $ is the strong coupling limit of
the $ E_{8} \times E_{8} $ heterotic string~\cite{HW}.

The low energy effective action of Ho\v{r}ava-Witten M-theory can be
formulated as an expansion in powers of the 11-dimensional gravitational
coupling $ \kappa $.  To lowest order in this expansion,
Ho\v{r}ava-Witten M-theory is 11-dimensional supergravity~\cite{CJS1},
which is of order $ \kappa^{-2} $.
The supergravity vacuum is specified by the metric $ g_{IJ} $, the 3-form
potential $ C_{IJK} $ with 4-form field strength 
$ G_{IJKL} = 24 \partial_{[I} C_{JKL]} $, and the spin $ 3/2 $ gravitino
$ \psi_{I} $. 

The $ \mathbb{Z}_{2} $ projection introduces gauge, gravitational, and
mixed anomalies into the theory.  Cancellation of the irreducible part of
the gravitational anomaly requires the introduction of two chiral 
$ \mathcal{N} = 1 $ $ E_{8} $ vector supermultiplets, one on each orbifold
fixed plane $ M^{10}_{i} $ $ (i=1,2) $ at $ x^{11} = 0 $ and 
$ x^{11} = \pi \rho $, respectively.  The reducible portion of the
gravitational anomaly, as well as the gauge and mixed anomalies, are
cancelled with a refinement~\cite{HW} of the standard
\emph{Green-Schwarz mechanism}~\cite{GS1}.  Requiring the one-loop chiral
anomaly to cancel fixes a relation between $ \kappa $ and the
10-dimensional gauge coupling $ \lambda $:
\begin{equation}
\frac{1}{\lambda^{2}} = \frac{1}{2 \pi \kappa^{2}} 
                        \left( \frac{\kappa}{4 \pi} \right)^{2/3}.
\end{equation}   
Formally, the low energy effective action of Ho\v{r}ava-Witten
M-theory appears to be an expansion with the $ m^{\textrm{th}} $
term being of order $ \kappa^{-2 + (2m/3)} $ $(m = 0,1,2...) $.  Other
exponents must arise at the quantum level since we will run into
infinities which, when cut off in the quantum theory, must on dimensional
grounds give anomalous powers of $ \kappa $.

\subsection{\label{Comp}Compactification to four dimensions}

We refer to the compactification of Ho\v{r}ava-Witten M-theory to lower
dimensions as \emph{heterotic M-theory}. The compactification to four
dimensions with unbroken $ \mathcal{N} = 1 $ supersymmetry was discussed
in~\cite{Witten;Strong}.  The procedure starts with the spacetime
structure
\begin{equation}
\label{spacetime}
M^{11} = M^{4} \times Z \times S^{1}/\mathbb{Z}_{2},
\end{equation}
where $ M^{4} $ is 4-dimensional Minkowski space and $ Z $ is a Calabi-Yau
3-fold.  M5-branes can be included in the bulk space at points throughout
the orbifold interval.  These M5-branes are required to span $ M^{4} $
(to preserve $ (3 + 1) $-dimensional Poincar\'{e} invariance) and wrap
holomorphic curves in $ Z $ (to preserve $ \mathcal{N} = 1 $ supersymmetry
in four dimensions).  

Generally, some subgroup $ G $ of the $ E_{8} $ symmetry will survive this
compactification.  $ E_{8} $ is broken to $ G \times H $, where the grand
unification group $ H $ is the commutant subgroup of $ G $ in $ E_{8} $.
The gauge fields associated with $ G $ `live' on the Calabi-Yau 3-fold,
and hence $ (3+1) $ Poincar\'{e} invariance is left unbroken.  The
requirement of unbroken $ \mathcal{N} = 1 $ supersymmetry implies that the
corresponding field strengths must satisfy the \emph{Hermitian Yang-Mills
constraints} $ F_{ab} = F_{\overline{a} \overline{b}} = g^{a
\overline{b}} F_{a \overline{b}} = 0 $.  Donaldson~\cite{Donald} and
Uhlenbeck and Yau~\cite{UY} prove that each solution to the
6-dimensional \emph{Hermitian Yang-Mills equations} $ D^{A}F_{AB} = 0 $
satisfying the Hermitian Yang-Mills constraints corresponds
to a semistable holomorphic vector bundle over the Calabi-Yau 3-fold with   
structure group being the complexification $ G_{\mathbb{C}} $ of the group
$ G $, and conversely.

The correction to the background~(\ref{spacetime}) is computed
perturbatively. The set of equations to be solved consists of the Killing 
spinor equation
\begin{equation}
\label{Killing}
\delta \psi_{I} = D_{I} \eta + \frac{ \sqrt{2} }{288}
   (\Gamma_{IJKLM} - 8 g_{IJ} \Gamma_{KLM}) G^{JKLM} \eta = 0,
\end{equation}
the equation motion
\begin{equation}
\label{motion}
D_{I} G^{IJKL} = 0
\end{equation}
and the Bianchi identity
\begin{multline}
\label{modBianchi}
(dG)_{11RSTU} = 4 \sqrt{2} \pi \left( \frac{\kappa}{4 \pi} \right)^{2/3}
  \Biggl[ J^{(0)} \delta(x^{11}) + J^{(N+1)} \delta (x^{11} - \pi \rho) 
\Biggr.           \\ 
\Biggl.  + \frac{1}{2} \sum_{n=1}^{N} J^{(n)}  
           \left( \delta (x^{11} - x_{n}) + \delta (x^{11} + x_{n})
           \right)   
  \Biggr]_{RSTU}
\end{multline}  
where $ J^{(0)} $, $ J^{(N+1)} $ are the sources on the orbifold fixed
planes at $ x^{11} = 0 $ and $ x^{11} = \pi \rho $, respectively, and 
$ J^{(n)} $ $ (n = 1, \ldots, N) $ are the M5-brane sources located at 
$ x^{11} = x_{1}, \ldots, x_{N} $ $ ( 0 \leq x_{1} \leq \ldots \leq x_{N}
\leq \pi \rho ) $.  Note that each M5-brane at $ x = x_{n} $ has to be
paired with a mirror M5-brane at $ x = - x_{n} $ with the same source
since the Bianchi identity must be even under the $ \mathbb{Z}_{2} $
symmetry.

The Bianchi identity~(\ref{modBianchi}) can be viewed as an expansion in
powers of $ \kappa^{2/3} $.  To linear order in $ \kappa^{2/3} $, 
the solution to the Killing spinor equation, equation of motion, and
Bianchi identity takes the form
\begin{gather}
(ds)^{2}  = (1 + b)\eta_{\mu \nu} dx^{\mu} dx^{\nu}
            + (g^{(CY)}_{AB} + h_{AB}) dx^{A} dx^{B}
            + (1 + \gamma) \left( dx^{11} \right)^{2}    \\
\begin{align}
G_{ABCD}  &= G^{(1)}_{ABCD}                              \\
G_{ABC11} &= G^{(1)}_{ABC11}                             \\
\eta      &= (1 + \psi) \eta^{(CY)}.
\end{align}
\end{gather}
with all other components of $ G_{IJKL} $ vanishing.  $ g^{(CY)}_{AB} $
and $ \eta^{(CY)} $ are the Ricci-flat metric and the covariantly constant
spinor on the Calabi-Yau 3-fold.  

As discussed in~\cite{LOW:4D}, the first order corrections 
$ b $, $ h_{AB} $, $ \gamma $, $ G^{(1)} $ and $ \psi $ can be expressed in 
terms of a single $ (1,1) $-form $ \mathcal{B}_{a \overline{b}} $ on the 
Calabi-Yau 3-fold.  All that remains then is to determine $ \mathcal{B}_{a
\overline{b}} $, which can be expanded in terms of eigenmodes of the
Laplacian on the Calabi-Yau 3-fold.  For the purpose of computing low
energy effective actions, it is sufficient to keep only the
zero-eigenvalue or `massless' terms in this expansion; that is, the terms
proportional to the harmonic $ (1,1) $ forms of the Calabi-Yau 3-fold. 
Let us choose a basis $ \{ \omega_{ia \overline{b}} \} $ for these
harmonic $ (1,1) $-forms, where $ i = 1, \ldots, h^{(1,1)} $.
We then write
\begin{equation}
\mathcal{B}_{a \overline{b}} = \sum_{i} b_{i} \omega^{i}_{a \overline{b}}
         + (\textrm{massive terms}).
\end{equation}
The $ \omega_{ia \overline{b}} $ are Poincar\'{e} dual to the 4-cycles 
$ \mathcal{C}_{4i} $, and one can define the integer charges
\begin{equation}
\beta^{(n)}_{i} = \int_{\mathcal{C}_{4i}} J^{(n)}, 
\quad n = 0,1,\ldots,N, N+1.
\end{equation}
$ \beta^{(0)}_{i} $ and $ \beta^{(N+1)}_{i} $ are the instanton charges on
the orbifold fixed planes and $ \beta^{(n)}_{i} $, $ n = 1, \ldots, N $
are the the magnetic charges of the M5-branes.  The expansion coefficients
$ b_{i} $ are found in~\cite{LOW:NS} in terms of these charges,
the normalized orbifold coordinates
\begin{equation}
\label{normorb}
z = \frac{x^{11}}{\pi \rho},  \quad z_{n} = \frac{x_{n}}{\pi \rho} \quad
(n = 1, \ldots N), \quad z_{0} = 0, \quad z_{1} = 1 
\end{equation}
and the expansion parameter
\begin{equation}
\epsilon = \left( \frac{\kappa}{4 \pi} \right)^{2/3} \frac{2 \pi^{2}
\rho}{\mathcal{V}^{2/3}},
\end{equation}
where $ \mathcal{V} = \int_{ Z } d^{6}x  \sqrt{ g^{(CY)} } $ is the
Calabi-Yau volume.

Finally, we note that a cohomological constraint on the Calabi-Yau 3-fold,
the gauge bundles, and the M5-branes can be found by integrating the
Bianchi identity over a 5-cycle which spans the orbifold interval together
with an arbitrary 4-cycle $ \mathcal{C}_{4} $ in the Calabi-Yau 3-fold.
Since $ dG $ is exact and the cycle is compact, this integral must vanish 
and we obtain
\begin{equation}
[W_{Z}] = c_{2}(TZ) - c_{2}(V_{Z1}) - c_{2}(V_{Z2})
\end{equation}
where $ c_{2}(TZ) $ and $ c_{2}(V_{Zi}) $ are the second Chern classes of
the tangent bundle $ TZ $ and the vector bundle $ V_{Zi} $, respectively  
and $ [W_{Z}] $ is the 4-form cohomology class associated with the  
M5-branes.

\subsection{Four-dimensional low energy effective theory}

Following~\cite{LOW:Five}, we now discuss the 4-dimensional low energy
effective theory on the observable orbifold fixed plane at $ x^{11} = 0 $.
As discussed in Section~\ref{Comp}, the apriori $ E_{8} $ gauge symmetry
is broken to $ G \times H $.  The $ \mathbf{248} $ of $ E_{8} $ decomposes 
under $ G \times H $ as 
$ \mathbf{248}_{E_{8}} \rightarrow \oplus_{\mathcal{S},\mathcal{R}}
(\mathcal{S},\mathcal{R}) $, where $ \mathcal{S} $ and $ \mathcal{R} $ are
irreducible representations of $ G $ and $ H $, with representation
indices $ x,y,\ldots = 1,\ldots,\textrm{dim}(\mathcal{S}) $ and
$ p,q,\ldots = 1,\ldots,\textrm{dim}(\mathcal{R}) $, respectively.
We denote a physical field in the representation $ \mathcal{R} $ of $ H $
by $ C^{Ip}(\mathcal{R}) $. Here $ I,J,K,\ldots = 1,\ldots, \textrm{dim} 
\left( H^{1}( Z,V_{ Z1 \mathcal{S} } ) \right) $ is the generation index,
$ V_{ Z1 \mathcal{S} } $ is the vector bundle $ V_{Z1} $ in the
representation $ \mathcal{S} $, and the cohomology group 
$ H^{1}(Z,V_{ Z1 \mathcal{S} }) $ has basis $ \{ u^{x}_{I} \} $.

Define the conventional 4-dimensional chiral fields $ S, T^{i} $ and the
chiral fields $ Z_{n} $ by
\begin{equation}
\textrm{Re}(S) = V; \quad \textrm{Re}(T^{i}) = Ra^{i}; \quad
\textrm{Re}(Z_{n}) = z_{n}
\end{equation}
Here $ V = \mathcal{V} / v $ where $ v = \int_{Z} d^{6}x $.
The nonvanishing components of the
Calabi-Yau metric are given by 
$ g^{(CY)}_{a \overline{b}} = g^{(CY)}_{\overline{b} a} = i a^{i}
\omega_{i a \overline{b}} $
$ ( i=1,\ldots,h^{(1,1)} ) $, where $ a^{i} $ are the $ (1,1) $ moduli of
the Calabi-Yau 3-fold.  The modulus $ R $ 
is the Calabi-Yau averaged orbifold radius divided by $ \rho $, and
the moduli $ z_{n} $ $ ( n = 0,1,\ldots,N,N+1 ) $ are
given by~(\ref{normorb}).   

The 4-dimensional low energy effective theory on the observable orbifold
fixed plane is specified in terms of 3 functions of the chiral matter
multiplets:

\begin{enumerate}

\item The \emph{K\"{a}hler potential} 
$ K_{matter} = Z_{IJ} \overline{C}^{I} C^{J} $ determines the kinetic
terms of the chiral matter fields.  To first order in the expansion
parameter $ \epsilon $, the K\"{a}hler metric $ Z_{IJ} $ takes the form
\begin{equation}
\label{Kmetric}
Z_{IJ} = e^{-K_{T}/3} \left[ G_{IJ} - \frac{\epsilon}{2 \mathcal{V}}
\tilde{\Gamma}^{i}_{IJ} \sum_{n=0}^{N+1} (1 - z_{n})^{2} \beta_{i}^{(n)}
\right],
\end{equation}
where 
\begin{align}
G^{ ( \mathcal{R} ) }_{IJ} &= \frac{1}{ \mathcal{V} } \int_{Z} 
  \sqrt{ g^{(CY)} } g^{ (CY) a \overline{b} } 
  u_{Iax}(\mathcal{R}) u^{x}_{J \overline{b}}(\mathcal{R})   \\
\tilde{\Gamma}^{i}_{IJ} 
 &= \Gamma^{i}_{IJ} - (T^{i} + \overline{T}^{i}) G_{IJ}
   -\frac{2}{3}( T^{i} + \overline{T}^{i} ) 
               ( T^{k} + \overline{T}^{k} ) K_{Tkj} \Gamma^{j}_{IJ}  \\
K_{T} &= -\textrm{ln} \left[ \frac{1}{6} d_{ijk} 
         ( T^{i} + \overline{T}^{i} )
         ( T^{j} + \overline{T}^{j} )
         ( T^{k} + \overline{T}^{k} ) \right]   
\end{align}
and
\begin{equation}
K_{Tij} = \frac{ \partial^{2} K_{T} }
                  { \partial T^{i} \partial \overline{T}^{j} }; \quad
\Gamma^{i}_{IJ} =  K^{ij}_{T} \frac{ \partial G_{IJ} }{ \partial T^{j} };
\quad
d_{ijk} = \int_{Z} \omega_{i} \wedge \omega_{j} \wedge \omega_{k}.
\end{equation}

\item The holomorphic \emph{superpotential} $ W $ determines the Yukawa
couplings
\begin{equation}
\label{Yukawa}
Y^{ ( \mathcal{R}_{1} \mathcal{R}_{2} \mathcal{R}_{3} ) }_{IJK}
 = 2 \sqrt{ 2 \pi \alpha_{G} }
   \int_{Z} \Omega \wedge u^{x}_{I}(\mathcal{R}_{1}) \wedge 
                          u^{y}_{J}(\mathcal{R}_{2}) \wedge
                          u^{z}_{K}(\mathcal{R}_{3}) 
            f^{ (\mathcal{R}_{1} \mathcal{R}_{2} \mathcal{R}_{3} ) }_{xyz}
\end{equation}
as well as the $ F $-term part of the scalar potential.  $ \Omega $ is the
covariantly constant $ (3,0) $ form and 
$ f^{ (\mathcal{R}_{1} \mathcal{R}_{2} \mathcal{R}_{3} ) }_{xyz} $
projects out the singlet in 
$ \mathcal{R}_{1} \times \mathcal{R}_{2} \times \mathcal{R}_{3} $ (if
any).  The Yukawa contribution to the superpotential is
\begin{equation}
W_{Y} = e^{K_{mod}/2} \frac{1}{3} Y_{IJK} C^{I}C^{J}C^{K},
\end{equation}
where $ K_{mod} = -\textrm{ln} (S + \overline{S}) + K_{T} $ is the moduli
contribution to the K\"{a}hler potential.
 
\item The holomorphic \emph{gauge kinetic function}
\begin{equation} 
f = S + \epsilon T^{i} \left[ \beta^{(0)}_{i} \sum_{n=1}^{N} 
                              (1 - Z_{n})^{2} \beta^{(n)}_{i} \right]
\end{equation}
determines the gauge kinetic terms and contributes to the gaugino masses
and the gauge part of the scalar potential.
    
\end{enumerate}
We note that~(\ref{Kmetric}) and~(\ref{Yukawa}) hold at the grand
unification scale $ M_{G} $, which coincides with the compactification
scale $ \mathcal{V}^{1/6} $.  The fermion mass hierarchies are encoded in
the K\"{a}hler metric, which must be diagonalized and rescaled to the unit
matrix to obtain the Yukawa couplings of the canonically normalized
fields~\cite{AD1}.  One then uses the supersymmetry renormalization group
equations to evaluate the Yukawa couplings at low energy.

If the perturbative correction to the background discussed in 
Section~\ref{Comp} is to make sense, the second term
in~(\ref{Kmetric}) must be a small correction to the
first.  However, setting $ \mathcal{V}^{1/6} = {M_{G}} = 3 \times 10^{16}
$ GeV, one finds $ \epsilon \simeq 0.93 $. Furthermore, one expects 
$ G_{IJ} $ and $ \tilde{\Gamma}_{IJ} $ to be of order 1.  Arnowitt and
Dutta~\cite{AD1} point out that the second term can still be a small
correction to the first if the instanton charges on the observable
orbifold fixed plane (at $ x^{11} = 0 $) vanish and the M5-branes cluster
near the hidden orbifold fixed plane (at $ x^{11} = \pi \rho $):
\begin{align}
\beta^{(0)}_{i} &= 0       \\
d_{n}           &\equiv (1 - z_{n}) \ll 1, \quad n = 1, \ldots, N.
\end{align}
We will impose the $ \beta^{(0)}_{i} = 0 $ constraint in
Section~\ref{Rules}.

\section{\label{Rules}Summary of rules}

In this section, we present rules for constructing a class of
$ \mathcal{N} = 1 $ supersymmetric nonperturbative vacua of
Ho\v{r}ava-Witten M-theory compactified on a
torus fibered Calabi-Yau 3-fold $ Z $ with first homotopy group 
$ \pi_{1}(Z) = \mathbb{Z}_{2} $, having 1) grand unification group 
$ H = E_{6} $, $ SO(10) $, or $ SU(5) $, 2) arbitrary net number of
generations $ N_{gen} $ of chiral fermions in the observable sector, and 3)
potentially viable matter Yukawa couplings.  The vacua with 
$ H = E_{6} $, $ SO(10) $, or $ SU(5) $ correspond to semistable
holomorphic vector bundles $ V_{Z} $ over $ Z $ having structure group $
G_{\mathbb{C}} = SU(n)_{\mathbb{C}} $ with $ n = 3 $, 4 or 5,
respectively, and generically contain M5-branes in the bulk space. 

\vskip 10pt
\noindent
{\bf Construction of $ Z $:}  We wish to construct a smooth \emph{torus}
fibered Calabi-Yau 3-fold $ Z $ with $ \pi_{1}(Z) = \mathbb{Z}_{2} $.  To
do this, we first construct a smooth \emph{elliptically} fibered
Calabi-Yau 3-fold $ X $ which admits a freely-acting involution $ \tau_{X}
$.  We can then construct the quotient manifold $ Z = X/\tau_{X} $.

\begin{itemize}

\item {\bf Construction of $ X $:}  To construct a smooth elliptically
fibered Calabi-Yau 3-fold $ X $ which admits a freely-acting involution 
$ \tau_{X} $,

\begin{enumerate}

\item {\bf Choose the base $ B $:}  The requirement that 
$ c_{1}(TX) = 0 $ restricts the possible bases~\cite{Grassi1,MV1}.  If the
base is smooth and preserves only $ \mathcal{N} = 1 $ supersymmetry in
four dimensions, then $ B $ is restricted to be a del Pezzo 
$ (dP_{r} $, $ r=0,1,\ldots,8) $, Hirzebruch $ (F_{r} $, $ r\geq 0) $,
blown-up Hirzebruch, or an Enriques surface $ (\mathcal{E}) $.

\item {\bf Require two global sections:}  To admit a freely-acting
involution $ \tau_{X} $, require $ X $ to have \emph{two} global sections
$ \sigma $ and $ \xi $ satisfying
\begin{equation}
\label{order2}
\xi + \xi = \sigma.
\end{equation}

Elliptically fibered manifolds can be described in terms of a Weierstrass
model.  A general elliptic curve can be embedded via a cubic equation into 
$ \mathbb{CP}^{2} $.  Without loss of generality, the equation can be
expressed in the Weierstrass form
\begin{equation}
\label{Wpoly}
z y^{2} = 4 x^{3} - g_{2} z^{2} x - g_{3} z^{3}
\end{equation}
where $ g_{2} $ and $ g_{3} $ are general coefficients and $ (x,y,z) $ are
homogeneous coordinates on $ \mathbb{CP}^{2} $.  To define an elliptic
fibration over a base $ B $, one needs to specify how the coefficients 
$ g_{2} $ and $ g_{3} $ vary as one moves around the base.  In order to
have a pair of sections $ \sigma $ and $ \xi $, the Weierstrass
polynomial~(\ref{Wpoly}) must factorize as
\begin{equation}
\label{Wpolyfac}
z y^{2} = 4 (x - az)(x^{2} + az x + bz^{2}).
\end{equation} 
Comparing~(\ref{Wpoly}) and~(\ref{Wpolyfac}), we see that
\begin{equation}
g_{2} = 4(a^{2} - b), \quad g_{3} = 4ab.
\end{equation}
The zero section $ \sigma $ is given by $ (x,y,z) = (0,1,0) $, and the
second section $ \xi $ by $ (x,y,z) = (a,0,1) $.

\item {\bf Blow up singularities:} The elliptic fibers are singular when
two roots of the Weierstrass polynomial~(\ref{Wpolyfac}) coincide.  The
set of points in the base over which the fibers are singular is given by
the discriminant locus
\begin{equation}
\Delta = 0
\end{equation}
where
\begin{equation}
\Delta = \Delta_{1} \Delta^{2}_{2}
\end{equation}
and
\begin{equation}
\Delta_{1} = a^{2} - 4b, \quad \Delta_{2} = 4(2a^{2} + b).
\end{equation}
One can show that there is a curve of singularities over the 
$ \Delta_{2} $ component of the discriminant curve.  To construct the
smooth Calabi-Yau 3-fold $ X $, one must blow up this curve of
singularities.  This is achieved by replacing the singular point of each
fiber over $ \Delta_{2} = 0 $ by a sphere $ \mathbb{CP}^{1} $.  This is a
new curve in the Calabi-Yau 3-fold, which we denote by $ N $.  The general
elliptic fiber $ F $ has now split into two spheres: the new fiber $ N $,
plus the proper transform of the singular fiber, which is in the class 
$ F - N $.   
 
\end{enumerate}

\item {\bf Choice of involution $ \tau_{X} $:}  Construct a freely-acting
involution $ \tau_{X} $ on $ X $ as the composition
\begin{equation}
\tau_{X} = \alpha \circ t_{\xi}
\end{equation}
where $ \alpha $ is the lift to $ X $ of a fibration-preserving involution
$ \tau_{B} $ on the base $ B $ with fixed point set 
$ \mathcal{F}_{\tau_{B}} $, and
\begin{equation}
t_{\xi}(x) = x + \xi(x), \quad x \in X
\end{equation}
is an involutive translation of the fibers.
To ensure that $ \tau_{B} $ preserves the fibration, require
\begin{equation}
\label{preserve}
\tau^{*}_{B}(a) = a, \quad \tau^{*}_{B}(b) = b.
\end{equation}
Upon the explicit specification of an involution $ \tau_{B} $ with the
above properties, the involution $ \alpha $ is uniquely determined by the
additional requirements that it fix the zero section $ \sigma $ and that
it preserve the holomorphic volume form on $ X $.

Note that $ \alpha $ leaves fixed the whole fiber above each point in 
$ \mathcal{F}_{\tau_{B}} $.  Since the action of translation on a
\emph{smooth} torus acts without fixed points, $ \tau_{X} $ will be freely
acting provided none of the fibers above $ \mathcal{F}_{\tau_{B}} $ are
singular.  Thus, require
\begin{equation}
\label{freeact}
\mathcal{F}_{\tau_{B}} \cap \{ \Delta = 0 \} = \emptyset.
\end{equation} 

\end{itemize}

\vskip 10pt
\noindent
{\bf Construction of a vector bundle $ V_{X} $ over $ X $ which descends
to a vector bundle $ V_{Z} $ over $ Z $:}

\begin{itemize}

\item {\bf $ G_{\mathbb{C}} = SU(n)_{\mathbb{C}} $ bundle constraints:}
We wish to construct (via the spectral cover
method~\cite{FMW,Donagi:Prin,BJPS1,DOPW}) a semi-stable holomorphic vector
bundle $ V_{X} $ over $ X $ with structure group $ G_{\mathbb{C}} =
SU(n)_{\mathbb{C}} $.  To do this, we need to fix a spectral cover $ C $
and a line bundle $ \mathcal{N} $ over it.  The condition that $
c_{1}(V_{X}) = 0 $ implies that the spectral data $ (C, \mathcal{N}) $ can
be written in terms of an effective divisor class $ \eta $ in the base $ B
$ and coefficients $ \lambda $ and $ \kappa_{i} $ $ (i = 1, \ldots, 4 \eta
\cdot c_{1}(B)) $.  Constraints are placed on $ \eta $ , $ \lambda $ , and
the $ \kappa_{i} $ by the condition that
\begin{equation}
c_{1}(\mathcal{N}) = n \left( \frac{1}{2} + \lambda \right) \sigma
   + \left( \frac{1}{2} - \lambda \right) \pi^{*}_{C} \eta 
   + \left( \frac{1}{2} + n \lambda \right) \pi^{*}_{C} c_{1}(B)
   + \textstyle{\sum_{i}} \kappa_{i} N_{i}
\end{equation}
be an integer class.  Various sufficient (but not necessary) constraints 
can be imposed~\cite{DOPW,DOPW2,FGI1}, but most generally, $
c_{1}(\mathcal{N}) $ will be an integer class if the constraints
\begin{gather}
q \equiv n \left( \frac{1}{2} + \lambda \right) \in \mathbb{Z}
\label{BCsigma} \\
\left(\frac{1}{2} - \lambda \right) \pi^{*}_{C} \eta
   + \left( \frac{1}{2} + n \lambda \right) \pi^{*}_{C} c_{1}(B)
\quad \textrm{is an integer class} \label{BCetaChern}   \\
\kappa_{i} - \frac{1}{2} m \in \mathbb{Z}, \quad m \in \mathbb{Z} 
\label{BCkappa}
\end{gather} are simultaneously satisfied. 

\item {\bf Bundle involution conditions:}  The bundle $ V_{X} $ over $ X $
will descend to a bundle $ V_{Z} $ over $ Z $ if $ V_{X} $ is invariant
under the involution $ \tau_{X} $. Necessary conditions for $ V_{X} $ to
be invariant are given by
\begin{align}
\tau_{B}(\eta) &= \eta    \label{BItau}\\
\textstyle{\sum_{i}} \kappa_{i} &= \eta \cdot c_{1}(B).   \label{BIkappa}
\end{align}
We note that there may be non-invariant bundles
satisfying~(\ref{BItau}) and (\ref{BIkappa}); the details of selecting
only the invariant bundles are beyond the scope of this paper.

\end{itemize}
{\bf Phenomenological constraints}

\begin{itemize}

\item {\bf $ N_{gen} $ condition:}  In the models of interest with 
$ V_{Z1} $ having structure group $ G_{\mathbb{C}} = SU(n)_{\mathbb{C}} $
(with $ n = 3 $, 4, or 5), the net number of generations 
($ \# $ generations $ - $ $ \# $ antigenerations) $ N_{gen} $ of chiral
fermions in the observable sector (in the $ \mathbf{27} -
\mathbf{\overline{27}}$ of $ E_{6} $, 
$ \mathbf{16} - \mathbf{\overline{16}} $ of $ SO(10) $, or $ \mathbf{ 10 +
\overline{5} } - ( \mathbf{ \overline{10} + 5 } ) $ of $ SU(5) $) is
given by 
\begin{equation}
N_{gen} = \frac{1}{2} \int_{Z} c_{3}(V_{Z1}).
\end{equation}
Since $ X $ is a double cover of $ Z $, it follows that
\begin{equation}
c_{3} (V_{Z}) = \frac{1}{2} c_{3}(V_{X}).
\end{equation}
$ c_{3}(V_{X}) $ has been computed by Curio~\cite{Curio1} and
Andreas~\cite{Andreas1}:
\begin{equation}
c_{3}(V_{X}) = 2 \lambda \sigma \wedge \eta \wedge ( \eta - n c_{1}(B) ).
\end{equation}
Thus, 
\begin{equation}
\label{Ngen}
N_{gen} = \frac{1}{2} \int_{B} \lambda \eta \wedge (\eta - n c_{1}(B))
        = \frac{1}{2} \lambda \eta \cdot ( \eta - n c_{1}(B) )
\end{equation}
where we have integrated over the fiber and used Poincar\'{e} duality.

\item {\bf Effectiveness condition:}  Anomaly cancellation requires
\begin{equation}
\label{W_{Z}}
[W_{Z}] = c_{2}(TZ) - c_{2}(V_{Z1}) - c_{2}(V_{Z2}),
\end{equation}
where $ [W_{Z}] $ is the class associated with non-perturbative M5-branes
in the bulk space of the theory.  For simplicity, we will take $ V_{Z2} $
to be the trivial bundle.  Hence, the gauge group $ E_{8} $ remains
unbroken in the hidden sector, $ c_{2}(V_{Z2}) $ vanishes,
and~(\ref{W_{Z}}) simplifies accordingly.  Condition~(\ref{W_{Z}}) can
then be pulled back onto $ X $ to give
\begin{equation}
\label{W_{X}}
[W_{X}] = c_{2}(TX) - c_{2}(V_{X1}).
\end{equation}   
The Chern classes appearing
in~(\ref{W_{X}}) have been evaluated to be~\cite{DOPW}
\begin{equation}
\label{c_2TX}
c_{2}(TX) = 12 \sigma_{*} c_{1} + \left( c_{2} + 11c^{2}_{1} \right)(F-N) 
            + \left( c_{2} - c^{2}_{1} \right) N 
\end{equation}
\begin{equation}
\label{c_2V_X}
c_{2}(V_{X}) = \sigma_{*} \eta - \left( f(n) - k^{2} \right) (F - N)
               - \left( f(n) - k^{2} + \textstyle{\sum_{i}} \kappa_{i}
                 \right) N
\end{equation}  
where $ c_{i} \equiv c_{i}(B) $ and
\begin{align}
k^{2} &= \textstyle{\sum_{i}} \kappa^{2}_{i} \\
f(n) &= \frac{1}{24} \left( n^{3} - n \right) c^{2}_{1}
            - \frac{1}{2} \left( \lambda^{2} - \frac{1}{4} \right)
              n \eta \cdot ( \eta - n c_{1} ).
\end{align}
Using these expressions for $ c_{2}(TX) $ and 
$ c_{2}(V_{X}) $,~(\ref{W_{X}})
becomes
\begin{equation}
[W_{X}] = \sigma_{*} W_{B} + c (F-N) + dN
\end{equation}
where
\begin{align}
\label{ceqn}
c &= c_{2} + f(n) + 11c^{2}_{1} - k^{2}     \\
\label{deqn}
d &= c_{2} + f(n) - c^{2}_{1} - k^{2} + \textstyle{\sum_{i}} \kappa_{i}    
\end{align}
and
\begin{equation}
W_{B} = 12c_{1}(B) - \eta. 
\end{equation}
The class  $ [W_{Z}] $ must represent a physical holomorphic curve in 
the Calabi-Yau 3-fold $ Z $ since M5-branes are required to wrap around
it. Hence $ [W_{Z}] $ must be an effective class, and its pull-back 
$ [W_{X}] $ is an effective class in the covering 3-fold $ X $.  Thus, we
require
\begin{equation}
W_{B} = 12 c_{1} - \eta  \quad \textrm{is effective in $ B $}
\end{equation}
and
\begin{equation}
c \geq 0, \quad  d \geq 0.
\end{equation}
 
\item {\bf $ \beta^{(0)}_{i} = 0 $ constraint:}  As discussed
in~\cite{AD1}, to obtain phenomenologically viable matter Yukawa
couplings, require vanishing instanton charges, $ \beta^{(0)}_{i} $, on
the observable orbifold fixed plane.  $ \beta^{(0)}_{i} = 0 $ implies that
\begin{equation}
\Omega \equiv c_{2}(V_{X1}) - \frac{1}{2} c_{2}(TX) = 0
\end{equation}
and thus from~(\ref{c_2TX}) and~(\ref{c_2V_X})
\begin{equation}
\sigma_{*} (6c_{1} - \eta) + \tilde{c} (F-N) + \tilde{d}N = 0
\end{equation}
where
\begin{align}
\tilde{c} &= c -\frac{1}{2} c_{2} - \frac{11}{2} c^{2}_{1}, 
\label{ctilde}   \\
\tilde{d} &= d -\frac{1}{2} c_{2} + \frac{1}{2} c^{2}_{1}.
\label{dtilde}
\end{align}
Thus, we require
\begin{equation}
\label{eta6c_1}
\eta = 6c_{1}(B) \\
\end{equation}
and
\begin{equation}
\quad \tilde{c} = 0 \quad \quad \tilde{d} = 0.
\end{equation}

\item{\bf Stability constraint:} Let $ G = SU(n) \subset E_{8} $ and
$ G_{\mathbb{C}} $ be the structure group of the vector bundle $ V_{Z} $.
Then the commutant subgroup of $ G $ in $ E_{8} $, denoted by $ H $ will be
the largest subgroup preserved by $ V_{Z} $ if~\cite{BM:Stab}
\begin{equation}
\eta \geq n c_{1}(B).
\end{equation}
We note that for the models of interest (which have $ n = 3 $, $ 4 $ and $
5 $), the $ \beta^{(0)}_{i} = 0 $ constraint~(\ref{eta6c_1}) ensures that
the stability constraint is satisfied.

\end{itemize}

\section{\label{Hirz}Hirzebruch surfaces}

In this section we demonstrate that torus-fibered Calabi-Yau manifolds 
$ Z $ with $ \pi_{1}(Z) = \mathbb{Z}_{2} $ and Hirzebruch base surfaces do
not admit the $ H = SO(10) $, $ N_{gen} = 3 $ vacua with potentially
viable matter Yukawa couplings that we seek.  

A Hirzebruch surface $ F_{r} $ $(r \geq 0)$, is a 2-dimensional complex
manifold constructed as a fibration with base $ \mathbb{CP}^{1} $ and
fiber $ \mathbb{CP}^{1} $.  We denote the class of the base and fiber of 
$ F_{r} $ by $ S $ and $ E $, respectively.  Their intersection numbers
are
\begin{equation}
S \cdot S = -r  \quad  S \cdot E = 1  \quad   E \cdot E = 0
\end{equation}
$ S $ and $ E $ form a basis of the homology class 
$ H_{2}(F_{r}, \mathbb{Z})$. This pair has the advantage that it is also
the set of generators for the Mori cone.  That is, the class
\begin{equation}
\eta = s S + e E
\end{equation}
is effective on $ F_{r} $ for integers $ s $ and $ e $ if and only if
\begin{equation}
s \geq 0,   \quad  e \geq 0.
\end{equation}
The Chern classes of $ F_{r} $ are
\begin{align}
c_{1}(F_{r}) &= 2 S + (r+2)E      \\
c_{2}(F_{r}) &= 4.   
\end{align}
We will need the result
\begin{equation}
c^{2}_{1}(F_{r}) = 8.
\end{equation}

\subsection{$ n=4 $ Hirzebruch solutions with $ N_{gen} = 3 $}
We wish to find $ n=4 $ Hirzebruch solutions,
corresponding to $ G=SU(4) $ and $ H=SO(10) $, with $ N_{gen} = 3 $.  We
begin by imposing the $ \beta^{(0)}_{i} = 0 $
constraint~(\ref{eta6c_1}):
\begin{equation}
\eta = 6c_{1}(F_{r}).
\end{equation}
With this constraint on $ \eta $, the second bundle involution
condition~(\ref{BIkappa}) becomes
\begin{equation}
\textstyle{\sum_{i}} \kappa_{i} = \eta \cdot c_{1}(F_{r}) = 6
c^{2}_{1}(F_{r}) = 48;
\quad i=1,\ldots,192
\end{equation} 
and the $ N_{gen} $ condition~(\ref{Ngen}), with $ N_{gen} = 3 $, becomes
\begin{equation}
3 = N_{gen} = \frac{1}{2} \lambda 6 (6 - n) c^{2}_{1}(F_{r}).
\end{equation}
For $ n=4 $, we obtain
\begin{equation}
\lambda = \frac{1}{16}.
\end{equation}
Plugging this value for $ \lambda $ along with $ n = 4 $
into~(\ref{BCsigma}), we see that the $ G_{\mathbb{C}} =
SU(4)_{\mathbb{C}} $ bundle constraints cannot be satisfied and hence
there are no $ n=4 $
Hirzebruch solutions with $ N_{gen} = 3 $ .  It is interesting to note
that the requirement of vanishing instanton charges on the observable
orbifold plane rules out the $ n = 4 $ Hirzebruch solutions presented
in~\cite{FGI1}.    

\section{\label{DPez}Del Pezzo surfaces}

In this section we demonstrate the existence of a torus fibered Calabi-Yau
3-fold $ Z $ with $ \pi_{1}(Z) = \mathbb{Z}_{2} $ and del Pezzo base
surface $ dP_{7} $ which admits $ H = SO(10) $, $ N_{gen} = 3 $ vacua
with potentially viable matter Yukawa couplings.

A del Pezzo surface $ dP_{r} $ $ (r = 0,1,\ldots,8) $, is a 2-dimensional
complex manifold constructed from complex projective space 
$ \mathbb{CP}^{2} $ by blowing up $ r $ points.  A basis of 
$ H_{2}(dP_{r}, \mathbb{Z}) $ composed of effective classes is given by
the hyperplane class $ l $ and $ r $ exceptional divisors $ E_{i} $, 
$ i = 1,\ldots,r $.  Their intersections are
\begin{equation}
l \cdot l = 1, \quad  E_{i} \cdot E_{j} = -\delta_{ij}, 
\quad E_{i} \cdot l = 0.
\end{equation} 
The Chern classes are given by
\begin{align}
c_{1}(dP_{r}) &= 3l - \sum_{i=1}^{r} E_{i}  \label{c_1dP_r} \\
c_{2}(dP_{r}) &= 3 + r.
\end{align}
We will need the result
\begin{equation}
c^{2}_{1}(dP_{r}) = 9 - r.
\end{equation}

\subsection{$ n = 4 $ del Pezzo solutions with $ N_{gen} = 3 $}
We wish to find $ n = 4 $ del Pezzo solutions, corresponding to 
$ G = SU(4) $ and $ H = SO(10) $, with $ N_{gen} = 3 $.  We begin by
imposing the $ \beta^{(0)}_{i} = 0 $ constraint~(\ref{eta6c_1}):
\begin{equation}
\label{eta6c_1dP_r}
\eta = 6c_{1}(dP_{r})
\end{equation}   
With this constraint on $ \eta $, the second bundle involution
condition~(\ref{BIkappa}) becomes
\begin{equation}
\textstyle{\sum_{i}} \kappa_{i} = \eta \cdot c_{1}(dP_{r}) = 6
c^{2}_{1}(dP_{r}) =
6(9-r); \quad i=1,\ldots,24(9-r)  
\end{equation}  
and the $ N_{gen} $ condition~(\ref{Ngen}), with $ N_{gen} = 3 $, becomes
\begin{equation}
3 = N_{gen} = \frac{1}{2} \lambda 6 (6 - n) c^{2}_{1}(dP_{r}).
\end{equation}
For $ n=4 $, we obtain
\begin{equation}
\label{lambdar}
\lambda = \frac{1}{2(9-r)}.
\end{equation}
The values of $ \lambda $ given by~(\ref{lambdar}) for each
$ r $ are given in Table~\ref{rlambda}.
\begin{table}
\label{rlambda}
\begin{tabular}{|c|c|c|c|c|c|c|c|c|c|}                  
\hline
 $ r $        & 0               & 1               & 2 
              & 3               & 4               & 5
              & 6               & 7               & 8             \\
\hline
 $ \lambda $  & 1/18            & 1/16            & 1/14   
              & 1/12            & 1/10            & 1/8    
              & 1/6             & 1/4             & 1/2           \\
\hline
 $ q $        & 20/9            & 9/4             & 16/7
              & 7/3             & 12/5            & 5/2
              & 8/3             & 3               & 4             \\
\hline 
\end{tabular}
\caption{The second line contains the $ n=4 $ del Pezzo $ dP_{r} $ values
for $ \lambda $ given by~(\ref{lambdar}).
$ q \equiv n \left( \frac{1}{2} + \lambda \right) $ must be an integer
for the $ G_{\mathbb{C}} = SU(n)_{\mathbb{C}} $ bundle
constraint~(\ref{BCsigma}) to be satisfied.} 
\end{table}
In this table, the quantity $ q \equiv n \left( \frac{1}{2} +
\lambda \right) $ must be an integer for the $ G_{\mathbb{C}} =
SU(n)_{\mathbb{C}} $ bundle constraint~(\ref{BCsigma}) to be satisfied.
From the table, we see that when $ n=4 $, this constraint can be satisfied
only for $ r = 7 $ or $ r = 8 $.  Thus, we can exclude the $ dP_{r} $
$ (r=0,\ldots,6) $ surfaces from consideration. 

We now try to satisfy the second $ G_{\mathbb{C}} = SU(n)_{\mathbb{C}} $
bundle constraint~(\ref{BCetaChern}).  Using~(\ref{eta6c_1dP_r}) 
in~(\ref{BCetaChern}) gives
\begin{equation}
\label{BCetaChernYuk}
\left[ \frac{7}{2} + \lambda ( n - 6 ) \right] \pi^{*}_{C} c_{1}(dP_{r})
\quad 
\textrm{is an integer class}.  
\end{equation}
Thus,~(\ref{BCetaChernYuk}) is satisfied if
\begin{equation}
p \equiv \frac{7}{2} + \lambda ( n - 6 ) \in \mathbb{Z}.
\end{equation}
For the $ r = 7 $ and $ r = 8 $ del Pezzo surfaces, we find
\begin{align}
p( \lambda = 1/4, n=4 ) &= 3  \\
p( \lambda = 1/2, n=4 ) &= 5/2 \notin \mathbb{Z}
\end{align}
respectively.  Thus, the $ r = 8 $ del Pezzo surface is excluded, and the
only remaining possibility is
\begin{equation}
\label{lambdadP_7n=4}
\lambda(r=7, n=4) = 1/4.
\end{equation}
We note that this value of $ \lambda $ would not be permitted if the 
sufficient (but not necessary) $ G_{\mathbb{C}} = SU(n)_{\mathbb{C}} $ 
bundle constraints discussed in~\cite{FGI1} had been imposed. 
Using~(\ref{lambdadP_7n=4}) and $ \eta = 6 c_{1}(dP_{7}) $ in~(\ref{ceqn})
and~(\ref{deqn}) gives
\begin{align}
c &= 46 - k^{2}    \label{ceqnk}   \\
d &= 34 - k^{2}.   \label{deqnk} 
\end{align}
Imposing the effectiveness conditions $ c \geq 0 $ and $ d \geq 0 $ gives
\begin{equation}
\label{ksquaredmax}
k^{2} \equiv \textstyle{\sum_{i}} \kappa^{2}_{i} \leq 34.
\end{equation}
Furthermore, $ \eta = 6c_{1}(dP_{7}) $ implies that 
\begin{equation}
W_{B} = 12c_{1}(dP_{7}) - \eta = 6 c_{1}(dP_{7})
\end{equation}
which means that $ W_{B} $ is indeed effective.

Using $ c^{2}_{1}(dP_{7}) = 2 $, $ c_{2}(dP_{7}) = 10 $ and our 
results $ c = 46 - k^{2} $, $ d = 34 - k^{2} $,~(\ref{ctilde})
and~(\ref{dtilde}) become   
\begin{equation}
\tilde{c} = \tilde{d} = 30 - k^{2}. 
\end{equation}
Imposing the $ \beta^{(0)}_{i} = 0 $ constraints 
$ \tilde{c} = \tilde{d} = 0 $ gives
\begin{equation}
\label{ksquareddP_7}
k^{2} \equiv \textstyle{\sum_{i}} \kappa^{2}_{i} = 30
\end{equation}
which is consistent with~(\ref{ksquaredmax}).  One needs therefore a set
of $ \kappa_{i} $ which simultaneously satisfy
\begin{equation}
\textstyle{\sum_{i}} \kappa_{i}  = 6 c^{2}_{1}(dP_{7}) = 12; \quad i =
1,\ldots,48
\end{equation}
and~(\ref{ksquareddP_7}) for $ \kappa_{i} $ obeying the bundle
constraint~(\ref{BCkappa}).  An example of such $ \kappa_{i} $ is
\begin{equation}
\kappa_{1} = \kappa_{2} = \kappa_{3} = 2, 
\quad \kappa_{4} = \kappa_{5} = 3,
\quad \textrm{all other} \ \kappa_{i} = 0.  
\end{equation}
Thus, $ n = 4 $ $ dP_{7} $ solutions with $ N_{gen} = 3 $  exist
whenever the constraints ~(\ref{preserve}),~(\ref{freeact}),
and~(\ref{BItau}) on the involution $ \tau_{X} $ are satisfied.

\section{\label{FreeFerm}Toward nonperturbative top quark mass}

In this section we discuss how the 11-dimensional framework of
Ho\v{r}ava-Witten M-theory may be used to extend the perturbative
calculation of the top quark Yukawa coupling in the realistic
free-fermionic models to the nonperturbative regime. 

Let us recall that in the free-fermionic heterotic string
formalism~\cite{ABK,KLT}, a model is
specified in terms of a set of boundary condition basis vectors and
one-loop GSO projection coefficients. The realistic free-fermionic
models of interest here are constructed in two stages.
The first stage corresponds to the NAHE set of boundary condition basis
vectors 
$ \{ \mathbf{1}, S, b_{1}, b_{2}, b_{3} \} $~\cite{FN1}.  At the second
stage, we add to the NAHE set three boundary condition
basis vectors, typically denoted by 
$ \{ \alpha, \beta, \gamma \} $. 
The gauge group at the level of the NAHE set is 
$ SO(6)^{3} \times SO(10) \times E_{8} $, 
which is broken to 
$ SO(4)^{3} \times U(1)^{3} \times SO(10) \times SO(16) $ by the vector 
$ 2 \gamma $.  Alternatively, we can start with an extended NAHE set 
$ \{ \mathbf{1}, S, \xi_{1}, \xi_{2}, b_{1}, b_{2} \} $ \cite{foc}, with 
$ \xi_{1} = \mathbf{1} + b_{1} + b_{2} + b_{3} $.  The set
$ \{ \mathbf{1}, S, \xi_{1}, \xi_{2} \} $ produces a toroidal Narain model
with $ SO(12) \times E_{8} \times E_{8} $ or 
$ SO(12) \times SO(16) \times SO(16) $ gauge group for appropriate choices
of the GSO phase $ c ( ^{\xi_{1}}_{\xi_{2}} ) $.  The basis vectors 
$ b_{1} $ and $ b_{2} $ then break $ SO(12) \rightarrow SO(4)^{3} $, and
either 
$ E_{8} \times E_{8} \rightarrow E_{6} \times U(1)^{2} \times E_{8} $
or
$ SO(16) \times SO(16) \rightarrow SO(10) \times U(1)^{3} \times SO(16) $. 
The vectors $ b_{1} $ and $ b_{2} $ correspond to $ \mathbb{Z}_{2} \times 
\mathbb{Z}_{2} $ orbifold modding.  The three vectors $ b_{1} $, $ b_{2}
$, and $ b_{3} $ correspond to the three twisted sectors of the 
$ \mathbb{Z}_{2} \times \mathbb{Z}_{2} $ orbifold, each producing eight
generations in the $ \mathbf{27} $ representation of $ E_{6} $ or 
$ \mathbf{16} $ representation of $ SO(10) $.  In the case of $ E_{6} $,
the untwisted sector produces an additional 
$ 3 \times ( \mathbf{27} + \mathbf{\overline{27}} ) $, whereas in the 
$ SO(10) $ model it produces 
$ 3 \times ( \mathbf{10} + \mathbf{\overline{10}} ) $.  Therefore, the
Calabi-Yau 3-fold which corresponds to the 
$ \mathbb{Z}_{2} \times \mathbb{Z}_{2} $ orbifold at the free-fermionic
point in the Narain moduli space has $ (h^{(1,1)}, h^{(2,1)}) = (27,3) $.  

This basic structure underlies all realistic free fermionic models.
In the second stage of the contruction the $SO(10)$ symmetry is broken
to one of its subgroups and the number of generations is reduced
to three, one from each of the twisted sectors $ b_{1} $, $ b_{2} $ or
$ b_{3} $. The top quark is identified with the leading mass state. The
Yukawa coupling of this mass state is obtained at the cubic level of the
superpotential and is a coupling between states from the  
twisted-twisted-untwisted sectors. For example, in the
standard-like models the relevant coupling is 
$ t_{1}^{c} Q_1{\bar h}_{1} $,
where $ t^c_{1} $ and $ Q_{1} $ are respectively the quark $ SU(2) $
singlet and doublet from the sector $ b_{1} $, and $ {\bar h}_1 $ is the
untwisted Higgs. Thus, one can calculate this coupling in the full three
generation model or at the level of the (51,3) or (27,3) $ \mathbb{Z}_2
\times \mathbb{Z}_2 $ orbifold, and as a
$ \mathbf{16}\cdot \mathbf{16}\cdot \mathbf{10} $ $ SO(10) $ coupling,
or a $ \mathbf{27}^{3} $ $ E_6 $ coupling. As long as the moduli are
fixed at the free-fermionic point, the numerical results will be   
identical. While we do not know the precise geometrical realization of the
three generation models, the geometry of the 
$ \mathbb{Z}_2\times \mathbb{Z}_2 $ orbifold is more readily identified.
Thus, to extend the calculation of the top quark Yukawa coupling in the
realistic free-fermionic models to the nonperturbative regime, one can
compactify Ho\v{r}ava-Witten M-theory on a Calabi-Yau 3-fold which
corresponds to the $ \mathbb{Z}_2\times \mathbb{Z}_2 $ orbifold.  
$ SU(n)_{\mathbb{C}} $ vector bundles with $ n = 3 $ or $ n = 4 $ can be
chosen, corresponding to the $ E_{6} $ or $ SO(10) $ grand unification
group, respectively.  The nonperturbative top quark Yukawa coupling at
the grand unification scale $ M_{G} $ is then computed, at least in
principle, using~(\ref{Yukawa}). However, this calculation may require
modifications to the rules presented in Section~\ref{Rules} in the sense
that we now discuss. 

Let $ X_{1} $ be the Calabi-Yau 3-fold which corresponds to the 
$ (51,3) $ $ \mathbb{Z}_{2} \times \mathbb{Z}_{2} $ orbifold.  As
discussed in~\cite{FGI1}, this manifold has the structure of the manifold
$ X $ described in Section~\ref{Rules}.
$ X_{1} $ can be realized as a singular limit of the $ (3,243) $
elliptically-fibered Calabi-Yau 3-fold
$ X^{'}_{1} $ with base
$ \mathbb{CP}^{1} \times \mathbb{CP}^{1} $~\cite{BEFNQ1}.  
We can represent the fibers of $ X^{'}_{1} $ in Weierstrass form
\begin{equation}
y^{2} = x^{3} + f_{8}(w,\tilde{w}) z^{4} x + g_{12}(w,\tilde{w}) z^{6}
\end{equation}
where $ w, \tilde{w} $ are inhomogeneous coordinates of the respective 
$ \mathbb{CP}^{1} $.  Making the choices
\begin{equation}
f_{8} = \eta - 3h^{2}, \quad g_{12} = h( \eta-2h^{2} ) 
\end{equation}
where
\begin{equation}
h = K \prod_{i,j=1}^{4} (w - w_{i}) (\tilde{w} - \tilde{w}_{j}), \quad 
\eta = C \prod_{i,j=1}^{4} (w - w_{i})^{2} (\tilde{w} -
      \tilde{w}_{j})^{2},
\end{equation}
we have a $ D_{4} $ singular fiber as we approach any of the $ w = w_{i} $
(or $ \tilde{w} = \tilde{w}_{j} $).  These $ D_{4} $ singularities
intersect in 16 points, $ (w_{i}, \tilde w_{j}) $, $ i,j = 1,\ldots,4 $ in
the base. To obtain the $ (51,3) $, resolving the singular fibers is not
enough.  One must also blow up the base once at each 
$ (w_{i},\tilde{w}_{j}) $, $ i,j=1,\ldots,4 $.  This blow-up procedure 
differs from the prescription in Section~\ref{Rules}.  Thus, a detailed
nonperturbative extension of the top quark Yukawa coupling calculation in
the realistic free fermionic models may require modifications to
the rules presented in Section~\ref{Rules}.  We remark that the
nonperturbative calculation of the remaining matter Yukawa
couplings~\cite{fm} requires more detailed knowledge of the geometry of
the three generation free-fermionic models.

\section{\label{Conc}Conclusions}

Using the rules presented in Section~\ref{Rules}, we have searched for 
$ \mathcal{N} = 1 $ supersymmetric nonperturbative vacua of
Ho\v{r}ava Witten M-theory compactified on a torus-fibered Calabi-Yau
3-fold $ Z $ with $ \pi_{1}(Z) = \mathbb{Z}_{2} $ having 
1) $ SO(10) $ grand unification group, 2) net number of generations 
$ N_{gen} = 3 $ of chiral fermions in the observable sector and 
3) potentially viable matter Yukawa couplings.  These vacua correspond to
semistable holomorphic vector bundles $ V_{Z} $ over $ Z $ having
structure group $ SU(4)_{\mathbb{C}} $, and generically contain M5-branes 
in the bulk space.  We have demonstrated that torus fibered Calabi-Yau
3-folds  $ Z $ with $ \pi_{1}(Z) = \mathbb{Z}_{2} $ and Hirzebruch base
surfaces do not admit such vacua, but those with a del Pezzo $ dP_{7} $
base surface do.  The extension of the top quark Yukawa coupling
calculation in the realistic free-fermionic models to the nonperturbative
regime was discussed.  It appears that a detailed analysis will require 
modifications to the rules presented in Section~\ref{Rules}.
We hope to make these modifications and perform a detailed analysis in a
future publication.  

\section*{Acknowledgements} 
We would like to thank Jose Isidro for discussions during the initial
stages of this project.  A.E.F. is supported in part by PPARC.


\end{document}